# Plasmonic superchiral lattice resonances in the mid-infrared

Francesco Mattioli[†,#], Giuseppe Mazzeo[‡,#], Giovanna Longhi[‡°], Sergio Abbate[‡°],
Giovanni Pellegrini[§], Michele Celebrano[§], Marco Finazzi[§], Lamberto Duò[§],
Chiara Zanchi[⊥], Matteo Tommasini[⊥], Marialilia Pea[†], Sara Cibella[†],
Filippo Sciortino[∥], Leonetta Baldassarre[∥], Alessandro Nucara[∥],
Michele Ortolani[∥], Paolo Biagioni[§,*]

† *Istituto di Fotonica e Nanotecnologie - CNR, Via Cineto Romano 42, 00156 Roma IT*

‡ *Dipartimento di Medicina Molecolare e Traslazionale, Università degli Studi di Brescia, Viale Europa 11, 25123 Brescia IT*

° *Istituto Nazionale di Ottica – CNR, Brescia Research Unit, c/o CSMT, via Branze 45, 25123 Brescia, IT*

§ *Dipartimento di Fisica, Politecnico di Milano, Piazza Leonardo da Vinci 32, 20133 Milano IT*

⊥ *Dipartimento di Chimica, Materiali e Ingegneria Chimica, Politecnico di Milano, Piazza Leonardo da Vinci 32, 20133 Milano IT*

∥ *Dipartimento di Fisica, Sapienza Università di Roma, Piazzale Aldo Moro 2, 000185 Roma IT*

\* E-mail: paolo.biagioni@polimi.it

# F. Mattioli and G. Mazzeo contributed equally

**Abstract**

Recent efforts in the field of surface-enhanced spectroscopies have focused on the paradigm of 'superchirality', entailing the engineering of the local electromagnetic fields to boost the enantiospecific interaction between light and chiral molecules. In this framework, approaches based on both metallic and dielectric nanostructures have been proposed and have also recently been extended to vibrational circular dichroism in the mid-infrared. In this work, we design, fabricate and characterize arrays of chiral plasmonic slits featuring enhanced chiral fields in the mid-infrared. We exploit collective lattice resonances to further enhance the local intensity and to generate sharp features in the circular dichroism spectra of the platform. Such features are ideally suited to test the superchiral coupling with the vibrational resonances of chiral molecules.

Since the seminal works by Tang and Cohen [Tang2010, Tang2011], several schemes have been developed by the nano-optics community to merge the sensitivity of surface-enhanced spectroscopies with the enantioselectivity provided by chiroptical techniques. The common goal of such 'superchiral' approaches is to engineer the local electromagnetic fields by means of metallic [Hendry2010, Govorov2010, Govorov2012, Abdulrahman2012, Frank2013, Lu2013, Valev2013, Valev2014, Schaferling2014, Tullius2015, Nesterov2016, Zhao2017, Garcia2018] or dielectric [Garcia2013, Pellegrini2017, Pellegrini2018, Mohammadi2018] nanostructures and enhance their ability to selectively interact with chiral molecules of a specific handedness. All such attempts stem from the consideration that, in the lowest-order dipolar approximation, the dissymmetry factor for the relative absorption between left and right circularly-polarized light by a molecule can be casted as $g = \left(\frac{G''}{\alpha''}\right)\left(\frac{2C}{\omega U_\mathrm{e}}\right)$, where $G''$ and $\alpha''$ are the imaginary part of the mixed and electric polarizabilities, respectively, $C = \frac{\varepsilon_0 \omega}{2}(\mathbf{E}^* \cdot \mathbf{B})$ is the 'optical chirality' and $U_\mathrm{e}$ is the energy density associated with the electric field. While in the original proposal the term 'superchiral' was specifically coined to define those field configurations that lead to a larger detected dissymmetry factor *g* compared to standard plane wave illumination [Tang2010], since then it has become common practice in the literature to employ the same term to define, more generally, also all those schemes that lead to an enhanced optical chirality *C*.

It can be fairly stated that the optimal requirements for surface-enhanced chiroptical spectroscopies are not yet fully established. When plasmonic enhancement is involved in the detection scheme, two resonances naturally come into play: the localized plasmon resonance in the nanoparticles and the electronic or vibrational resonance in the chiral molecules. Most of the experimental results published so far with localized plasmons are based on chiroptical refractometric sensing in the visible spectral range, i.e. on the different refractive index that characterizes non-absorbing chiral molecules when they are illuminated *off resonance* by fields with opposite optical chirality. The effect of such a contrast can be observed by means of circular dichroism (CD) spectra *on resonance* with the localized plasmons of the system. In this approach, both chiral and non-chiral plasmonic structures have been proposed and the molecular chirality typically manifests itself via the shift and/or the change in the lineshape induced on the plasmonic spectral features [Hendry 2010, Abdulrahman2012, Lu2013, Tullius2015, Zhao2017]. Recently, also racemic arrays of chiral plasmonic antennas, featuring both types of handedness in the same array, were successfully employed [Garcia2018].

Translating plasmon-enhanced chiroptical spectroscopies to the mid-infrared region allows addressing the vibrational resonances of molecules and thus increasing the sensitivity to their conformation and configuration [Knipper2018, Knipper2018b]. In this case, both the plasmonic platform and the molecules are illuminated *on resonance*, possibly giving rise to the vast zoology of interferential lineshapes that are generated by the two coupled oscillators and that are expected to manifest themselves in the CD spectra as well.

In this work we design, fabricate, and characterize a superchiral plasmonic platform in the mid-infrared. The individual building blocks are chiral coupled slits such as those sketched in **Figure 1a**, previously introduced by Hendry et al. [Hendry2012]. The two slits provide resonant enhancement of the optical chirality in the region of space right above the sample surface thanks to the favorable superposition of the respective electric and magnetic fields. This system is particularly appealing for use in transmission geometry, as customary for CD experiments, since it features an enhanced transmissivity in correspondence with the dipolar plasmonic resonance. However, the resonance is broad, with a quality factor of only about 5, which is typical for the dipolar resonances of linear gold antennas or slits operating in the mid-infrared. Moreover, the strong linear anisotropy of the slits gives rise to dramatic artifacts that overwhelm the genuine CD signal. In order to counterbalance these drawbacks, we design a resonant slit array where the superchiral response of the individual slit pairs is combined with a collective lattice resonance giving rise to: (i) enhanced superchiral near fields, (ii) narrower spectral features, and (iii) a squared arrangement that removes the linear anisotropy of the individual building blocks. In perspective, this platform is an ideal benchmark to assess the chiroptical coupling between plasmonic and molecular (vibrational) oscillators and it is also viable to an overall achiral arrangement such as the one introduced in Ref. [Garcia2018], which is especially appealing because it provides enhanced chiral near fields without any chiral background in the CD spectra due to the plasmonic platform.

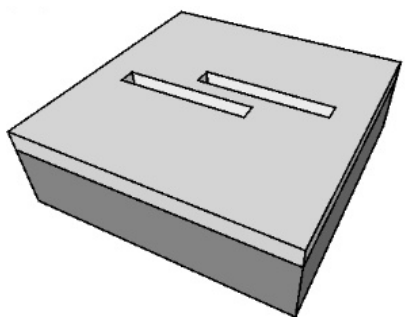

**Figure 1:** Sketch of the individual building block featuring two slits in a chiral configuration.

**Results and Discussion**

**Sample design and fabrication**

The coupled chiral slits that represent the individual building blocks for the array have already been introduced in the literature previously [Hendry2012]. Their specific asset is that the electric and magnetic fields associated with their fundamental electric dipole resonance superimpose in a favorable way in order to locally enhance the optical chirality. **Figure 2a** demonstrates the results of finite-difference time-domain (FDTD) simulations for a slit pair illuminated on resonance at 1300 cm$^{-1}$ wavenumbers (about 7.7 μm wavelength). The map plots the optical chirality on a plane located 10 nm above the sample surface, normalized to the optical chirality of a circularly polarized plane wave. A hot spot of enhanced optical chirality is indeed observed in the central part of the map.

The most investigated spectral region for VCD is roughly in the 900-1800 cm$^{-1}$ range [Nafie2011] and the investigated Au slits can easily be tuned to cover the whole window by simply changing their length. We fabricate a first set of calibration samples in which the thickness of the metal film (50 nm), the width of the two slits (200 nm), and the gap between them (150 nm) are kept fixed and only their length is varied. A representative scanning electron microscopy image of one of the samples is shown in **Figure 2b**. The slits are fabricated on a CaF$_2$ substrate using electron beam lithography (100 kV Vistec EBPG 5HR) and lift-off techniques. A negative electronic resist bilayer made of PMMA 669.06 (thickness 300 nm) and HSQ XR1541.006 (thickness 150 nm) is spun on top of the substrate and then covered with a thermally-evaporated Al layer (thickness 5 nm) to prevent charging effects during the exposure. After the exposure and the development, the slit geometry is defined on the HSQ layer and an oxygen-based reactive ion etching is used to remove the PMMA underneath and obtain the correct bilayer profile needed to perform a nanometric lift-off process. Finally, a 10 nm/40 nm Ti/Au layer is grown by electron-gun evaporation and then lifted off. The samples extend over areas of about 300×300 μm$^2$ and are characterized by means of Fourier-transform IR (FTIR) microscopy, employing an IR microscope based on a cassegrain objective in reflection geometry, with a linear electric field polarization perpendicular to the long axis of the slit. The experimental results in **Figure 2c** confirm the wide tunability of the fundamental localized plasmon resonance (observed as a dip in the reflected signal) and are reproduced with excellent agreement by simulations performed with the FDTD method [FDTD], as demonstrated in **Figure 2d**, which are performed under plane-wave illumination with a refractive index $n$=1.35 for the CaF$_2$ substrate and the Au dielectric constant taken from the literature [Olmon2012].

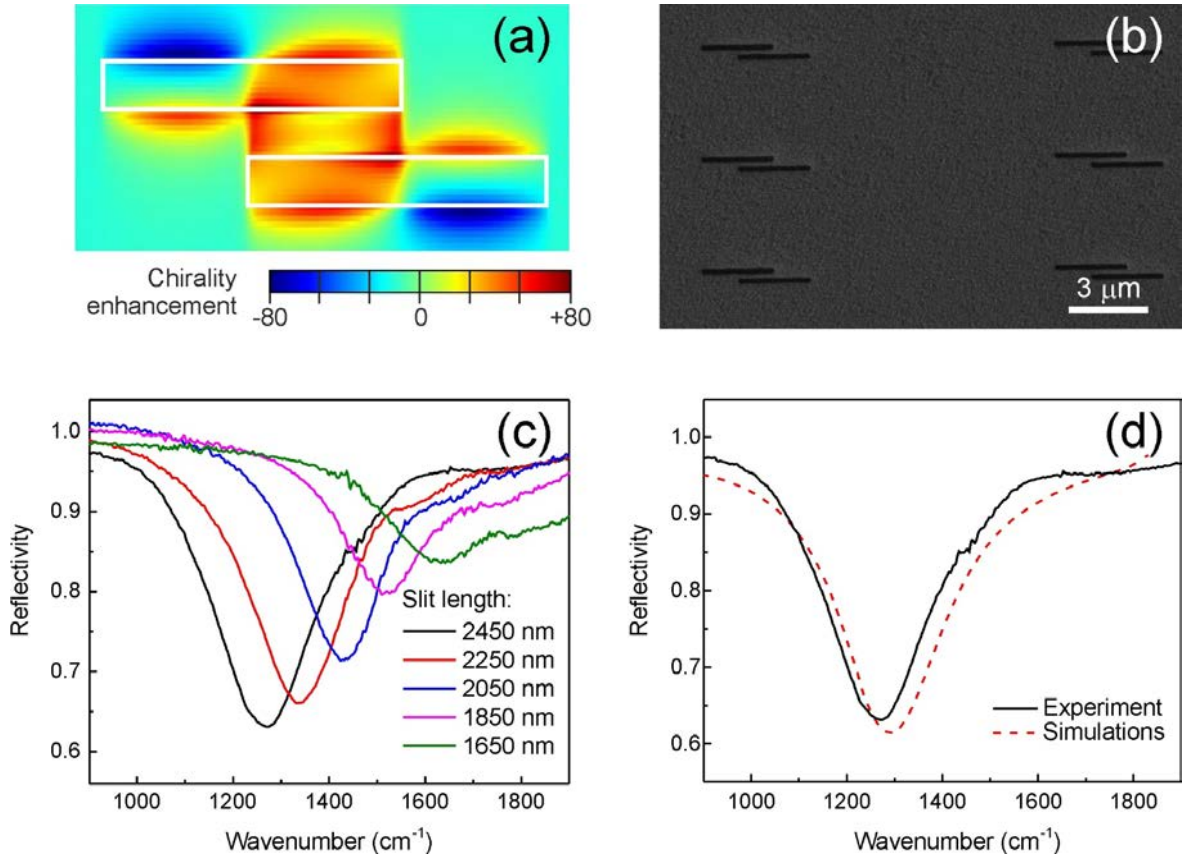

**Figure 2:** (a) FDTD simulation of the hot spot of optical chirality created by the coupled slits; (b) representative SEM image of the fabricated slits (the widths of the slits and the gap between them are 200 nm and 150 nm, respectively); (c) experimental reflectivity spectra acquired from slits of different length (all spectra are normalized to the reflectivity of a gold mirror); (d) representative comparison between the experimental and the simulated reflectivity spectra for a length of 2450 nm of the individual slits.

However, when it comes to CD measurements, this design has the disadvantage of a strong linear optical anisotropy, which introduces dramatic well-known artifacts in the measurement of a CD spectrum. In particular, any slight misalignment of the optical path (typically, a slightly off-normal incidence) combined with the presence of a uniaxial anisotropy in the plane of the sample surface creates the conditions for what has also been referred to as 'extrinsic chirality' in the literature, an effect that can easily overwhelm any genuine CD signal [Drezet2008, Plum2009, Sersic2012].

For this reason, we design a modified sample geometry in which the chiral slit pairs are arranged in a square lattice, as shown in **Figure 3a**. By doing this, we also take advantage of the closely packed arrangement to tune the lattice resonances in the system so they coherently excite the localized resonances of the nanostructures in the array. This spectral overlap is accompanied by strong local near fields, which have e.g. already been exploited in the literature for sensing or emission enhancement [Adato2009, Giannini2010], and by discontinuities in the reflection/transmission spectra (associated with the so-called Wood-Rayleigh anomalies) that create sharp spectral features. We fabricated a large-area sample of 5×5 mm$^2$ with a square array with a 6 μm pitch and a slit length of 2.5 μm and we characterized it by FTIR spectroscopy. With our collection geometry in the forward direction (standard transmission measurements) we probe the 0$^{th}$ diffraction order, while the periodicity of the square lattice sets the appearance of the 1$^{st}$ order in air around 1666 cm$^{-1}$ and of the 1$^{st}$ order in the CaF$_2$ substrate around 1235 cm$^{-1}$. Moreover, the diagonal of the square unit cell, which is about 8.5 μm, further contributes to the overall response by opening the possibility for a diffraction order in air around 1160 cm$^{-1}$. The transmission spectrum (**Figure 3b**), clearly reveals a more complex response than the one displayed by the uniaxial array, with narrower lineshapes and sharp features marking the discontinuities introduced by the diffraction orders. Noticeably, the experimental line widths are limited by the non-collimated thermal FTIR source, which, as expected, broadens any feature associated with grating effects [Baldassarre2013]. For sake of clarity, it should be noted that such transmission spectra are here expressed and labeled as 'absorbance', as customary for CD measurements, although no significant absorption is present for gold plasmonic nanostructures in the mid-infrared and light is mainly either reflected or transmitted by the system. To avoid converge problems that typically affect time-domain methods applied to structures endowed with sharp resonances, this sample is modeled by employing the rigorous coupled-wave analysis (RCWA), a frequency-domain method that is particularly suited for periodic systems [RETICOLO]. In order to mimic the ~2° half angle of the almost-collimated illumination in the FTIR setup, we run separate plane-wave simulations for different angles of incidence and we then sum them to construct the final transmission spectrum. The results, also shown in **Figure 3b**, display a fair agreement with the experiment.

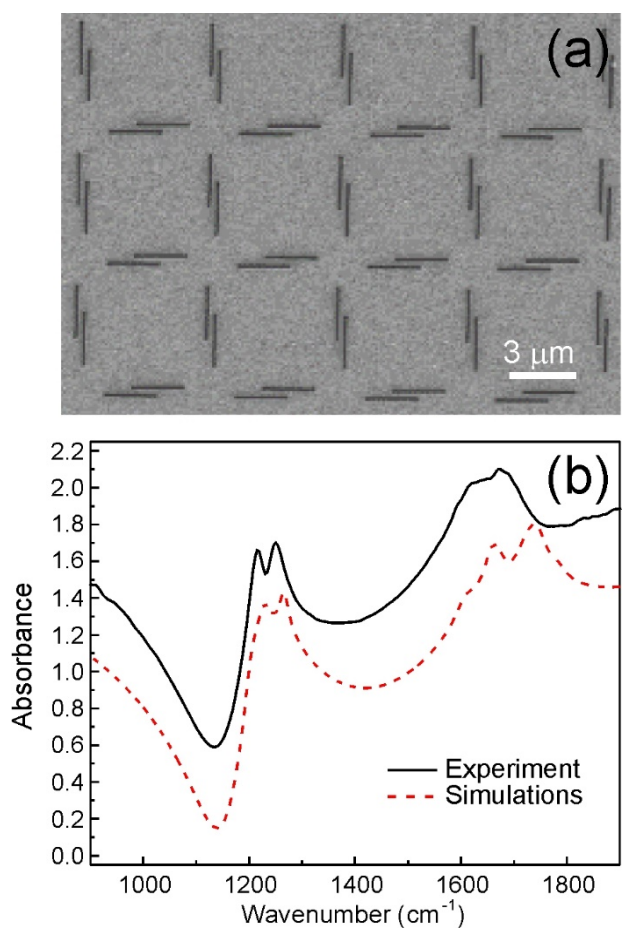

**Figure 3:** (a) Representative SEM image of the square slit array with 6 μm pitch (the widths of the slits and the gap between them are 200 nm and 150 nm, respectively, as in the previous case); (b) transmission FTIR spectrum acquired with a quasi-collimated source (incidence half angle about 2°) and the associated RCWA simulations obtained by averaging over the same range of angles.

**Chiroptical characterization**

The CD of the plasmonic array is investigated with a JASCO FVS6000 apparatus, which is equipped with a high intensity ceramic source, a MCT-V (HgCdTe) diode detector, and a ZnSe photo-elastic modulator for the lock-in acquisition of CD spectra. We study two samples characterized by opposite handedness of the individual slit pairs. To confirm that the measurements are reasonably free from artifacts, possibly induced by the planar two-dimensional geometry of the sample as discussed before, we acquire the VCD spectra as a function of the azimuthal angle. **Figure 4a** displays a set of measurements acquired for the two plasmonic enantiomers (black and red curves, respectively). The results confirm that (i) the sharp spectral features of the sample transmissivity translate into sharp CD lineshapes; (ii) the CD spectra from the two enantiomers are the mirror image of each other, as expected, and (iii) the measurements are reasonably free from artifacts, as demonstrated by the fair reproducibility of the spectra collected at different azimuthal angles. The presence of the sharp features in the CD spectrum, in particular, is appealing in view of sensing schemes in which the coupling between the plasmonic and the molecular vibrational oscillators is exploited and analyzed. It should be stressed that, for such two-dimensional chiral systems, it is expected that the CD spectrum is also reversed upon flipping the sample with respect to the direction of light propagation [Fedotov2007]. This is confirmed by the measurements in **Figure 4b**, in which the black spectra are the same as those in **Figure 4a** while the gray spectra are acquired from the same flipped plasmonic enantiomer.

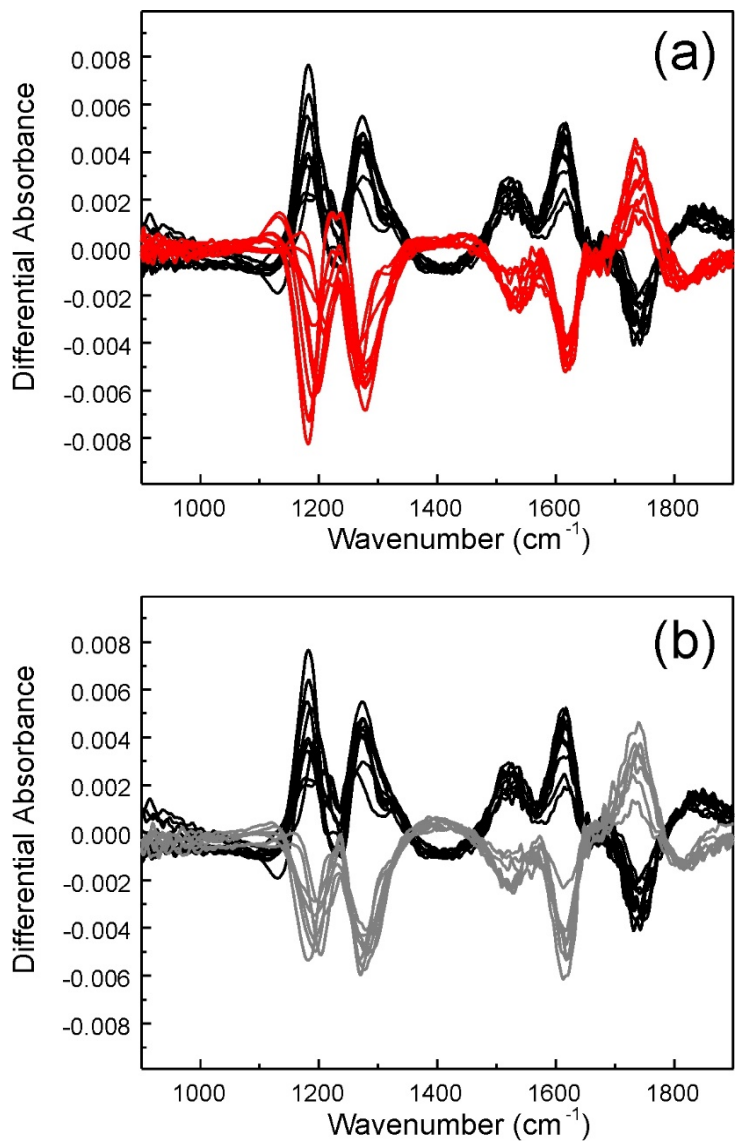

**Figure 4:** (a) CD spectra acquired for two opposite plasmonic enantiomers (black and red lines, respectively) for different azimuthal angles in steps of 20°; (b) CD spectra acquired for the same plasmonic enantiomer after flipping the sample with respect to the propagation direction of the excitation beam (black and grey lines, respectively, the black lines represent the same data as those in panel a) for different azimuthal angles in steps of 20°.

**Discussion and conclusions**

In conclusion, we have designed and characterized a periodic array of chiral slit pairs, featuring superchiral near fields whose intensity and spectral response is enriched by the presence of lattice resonances in the system, resulting in extremely narrow CD lineshapes. The plasmonic array represents an ideal system for surface-enhanced vibrational CD in transmission geometry: thanks to the resonantly enhanced transmissivity through the slits, only the molecules located in proximity of the superchiral hot spots are probed in the far field. Along this line, the array represents an ideal benchmark to test the chiroptical coupling between localized plasmon resonances, lattice resonances, and molecular vibrational resonances of chiral molecules. Moreover, the design is viable to be modified in order to achieve an overall achiral plasmonic response by suitably alternating left and right slit pairs, thus increasing the sensitivity to the molecular response without an overwhelming plasmonic response, as recently proposed in Ref. [Garcia2018].

**Acknowledgements**

P.B. acknowledges the kind support of Philippe Lalanne in the use of the RETICOLO software. The research leading to these results has received funding from the Italian Ministry of Education, University, and Research (MIUR), PRIN project 'Plasmon-enhanced vibrational circular dichroism', ID 2015FSHNCB.